# Definitive experimental evidence for two-band superconductivity in MgB$_2$

One-sentence summary: We provide definitive experimental evidence for the two-band superconductivity of MgB$_2$ as the origin of the multiple superconducting gap.

S. Tsuda,[1] T. Yokoya,[1] Y. Takano,[2] H. Kito,[3] A. Matsushita,[2] F. Yin,[2] H. Harima,[4] S. Shin[1,5]

[1]Institute for Solid State Physics, University of Tokyo, Kashiwa, Chiba 277-8581, Japan

[2]National Institute for Materials Science, 1-2-1 Sengen, Tsukuba 305-0047, Japan

[3]National Institute of Advanced Industrial Science and Technology, 1-1-1 Umezono, Tsukuba 305-8568, Japan

[4]The Institute of Scientific and Industrial Research, Osaka University, Ibaraki, Osaka 567-0047, Japan

[5]The Institute of Physical and Chemical Research (RIKEN), Sayo-gun, Hyogo 679-5143, Japan

The superconducting gap of MgB$_2$ has been studied by high-resolution angle-resolved photoemission spectroscopy (ARPES). The momentum($k$)-resolving capability of ARPES enables us to identify the σ- and π-orbital derived bands predicted from band structure calculations and to successfully measure the superconducting gap on each band. The results show that superconducting gaps with values of 5.5 meV and 2.2 meV open on the σ-band and the π-band, respectively, but both the gaps close at the bulk transition temperature, providing a definitive experimental evidence for the two-band superconductivity in MgB$_2$. The experiments validate the role of $k$-dependent electron-phonon coupling as the origin of multiple-gap superconductivity in MgB$_2$.

Bardeen-Cooper-Schrieffer (BCS) theory assumes that phonons mediate electron paring with a momentum ($k$)-independent electron-phonon coupling constant, giving rise to a simple isotropic superconducting energy gap (1). For $MgB_2$, with an unexpectedly high transition temperature ($T_c$) (2) among the phonon-mediated superconductors (3), however, evidence for a multiple gap, and not a simple isotropic gap, has been obtained from $k$-integrated probes (4). Therefore, the origin of the multiple gap and its relation to the mechanism yielding such a high $T_c$ are strongly debated (5-9). Recently, first principle theoretical studies with $k$-dependent electron-phonon coupling have shown that two gaps with different magnitudes open on boron $\sigma$- and $\pi$-orbital derived bands (10). However, possibilities of other scenarios predicting and/or explaining similar multiple-gap structures (8,9) can not be totally ruled out unless genuine $k$-resolved experimental evidence is obtained. Here, we show results of the superconducting gap of $MgB_2$ measured on different bands. The results clearly identify a larger gap on the $\sigma$-band and a smaller gap on the $\pi$-band, but which close at the same temperature ($T_c$), providing definitive experimental evidence for the two-band superconductivity in $MgB_2$. The present results together with the recent first principle band calculations (10) establishes the two-band superconductivity in $MgB_2$ and will motivate further experimental and theoretical studies for researching higher-$T_c$ materials based on the two-band superconductivity.

The crystal structure of $MgB_2$ consists of honeycomb boron layers separated by magnesium layers.[2] Band structure calculations have indicated that magnesium is substantially ionised, and the bands at the Fermi level ($E_F$) derived mainly from boron orbital with very different characters (11-13). As a result, $MgB_2$ possesses total four Fermi surface (FS) sheets that can be classified into two types; two 2-dimensional (2D) cylindrical FS sheets around the Γ-A line derived from σ-bonding states of the boron $p_{x,y}$ orbitals and two 3-dimensional (3D) tubular sheets derived from π- bonding and antibonding states of the boron $p_z$ orbitals. The intersection of the FS's with the $k_z = 0$ plane are shown in Fig. 1A, where two small (solid lines) and one large (broken line) circular sheets correspond to the 2D FS sheets and one of the 3D sheet (14). The size of the superconducting gap in relation to the σ- and π-bands (or FS's) can be studied by using angle-resolved photoemission spectroscopy (ARPES) that measures the $k$- and temperature-dependent electronic states of a solid. These capabilities of ARPES are unique and have been used for studying the $k$-dependent superconducting gap in high-temperature superconductors (15,16), and FS sheet-dependent superconducting gap in transition metal dichalcogenide 2$H$-NbSe$_2$ (17). However, in MgB$_2$, the limitation of obtained sizes for single crystal samples (typically several hundred micrometer) has prevented measuring superconducting gap due to the very small count when used with higher resolution, though medium-resolution ARPES study has reported the valence band dispersions of MgB$_2$ (18).

The single crystals of MgB$_2$ used in this study were prepared with high pressure synthesis by a similar technique as described earlier (19). Magnetization measurements confirmed that the samples have a middle point of the superconducting transition at 36 K. ARPES measurements were performed on a spectrometer built using a Scienta SES2002 electron analyser and a GAMMADATA high-flux discharging lamp with a toroidal grating monochromator using He IIα (40.814 eV) resonance lines. The energy and angular resolution for the valence band spectra were set to ~100 meV and ± 0.1 deg (corresponding to 0.010 Å$^{-1}$), respectively. The high-resolution measurements for the superconducting gap for the σ-band and the π-band were set to energy resolution of 3.9 and 5.0 meV, respectively, depending on the count rate. Samples are cooled using a flowing liquid He refrigerator with improved thermal shielding. Sample temperature was measured using a silicon-diode sensor mounted below the samples. The base pressure of the spectrometer was better than $3 \times 10^{-11}$ Torr. The sample orientation was checked by symmetry of ARPES spectra and further confirmed by electron diffraction studies after the ARPES measurements. All the ARPES measurements have been done for *in-situ* cleaved surfaces. Temperature-dependent spectral changes were confirmed by cycling temperature across $T_c$. $E_F$ of samples for high-resolution measurements was referenced to that of a gold film evaporated onto the sample substrate and its accuracy is estimated to be better than $\pm$ 0.2 meV.

Figure 1B shows an intensity plot of ARPES spectra of $MgB_2$ measured along the $\Sigma(R)$ high symmetry line (line 1 in Fig 1A), compared with the calculated band dispersions (14) on the $k_z = 0$ plane (solid lines for the $\sigma$-band and broken lines for the $\pi$-band). One can clearly see a prominent feature that disperses toward $E_F$ in the first Brillouin zone (BZ). Having a fairly good agreement with the calculated boron 2p $\pi$-orbital derived band in terms of the dispersion and the $k$ point where it crosses $E_F$ ($k_F$), this band can be ascribed to the $\pi$-band in the $k_z = 0$ plane. Observation of the $\pi$-band in the $k_z = 0$ plane suggests that measured $k_z$ position with the photon energy we used is located near the $k_z = 0$ plane. Besides the prominent structure, we also find another dispersive feature that crosses $E_F$ near the $\Gamma(A)$ point in the second BZ. It is found that this band follows the dispersion of one of the calculated bands derived from the boron 2p $\sigma$-orbital and can be assigned to the $\sigma$-band. Not being observed the $\pi$-band in the second BZ and the other $\sigma$-band in the second BZ can be attributed to matrix element effects, which strongly affects the intensity of photoelectrons as has been demonstrated for single crystal graphite (20). Figure 1C and 1D are the ARPES intensity plots measured along the lines 2 and 3 in Fig. 1A, respectively. In both measured directions, we see a band crossing and can obtain $k_F$'s. As shown in Fig. 1A, those $k_F$'s determined from the present ARPES (red open circles) are corresponding to the calculated $\sigma$- and $\pi$-orbital derived FS sheets and therefore each $k_F$ of the lines 2 and 3 can be attributed to that of the $\sigma$- and $\pi$-bands, respectively. The previous medium-resolution ARPES measurements (18) in the first BZ observed the σ- and π-bands with an additional surface derived band superimposed on the σ-band in the near $E_F$ region. In the present study, we found no evidence for the surface derived band in the second BZ facilitating the superconducting-gap measurements on the σ-band.

Thus, the identification of the both σ- and π-band derived FS sheets provides us the opportunity to measure the superconducting gap on each band (or each FS sheet). Figures 2A and 2B show temperature-dependent high-resolution ARPES spectra near $E_F$ obtained along the lines 2 and 3 in Fig. 1A, respectively. The temperature-dependent spectra measured for the σ-band show redistribution of spectral weight from the region near $E_F$ to higher binding energy as the temperature is lowered, indicating opening of superconducting gap below $T_c$. At 6 K, the spectrum has a peak around 10 meV and a leading-edge shift of 3.6 meV. The leading-edge shift of a superconducting spectrum has been used to qualitatively estimate the size of a superconducting gap (21). The spectra measured at the π-band also show similar temperature dependence. The spectrum measured at 6 K has a much pronounced condensation peak, but shows a leading-edge shift of only 0.9 meV. The larger shift on the σ-band than on the π-band in spite of the lower resolution used for the σ-band strongly indicates that sizes of the superconducting gap are highly dependent on the character of the bands. The small shoulder structure near $E_F$ of 30 K spectra of the π-band (inset of Fig. 2B) is indicative of persistence of the superconducting gap close to the bulk $T_c$ of 36 K, indicating that the smaller gap on the π-band originates in the bulk electronic structure, too. Therefore we can conclude that these raw data itself provide first definitive experimental evidence for the two-band superconductivity of $MgB_2$.

For quantitative study, we estimated the size of the gap using modified BCS function that includes the magnitude of a superconducting gap Δ and the quasiparticle lifetime broadening Γ (Ref. 22). The results are superimposed with lines on the energy-enlarged spectra shown in insets of Figs. 2A and 2B. It can be seen that the function can well reproduce the raw spectra especially for the gap for the $\pi$-band even for the small shoulder structure above $E_F$, which can be explained by thermally excited electrons over the gap (23). The size of the superconducting gap for the $\sigma$- and $\pi$-band are plotted as a function of temperature with open circles and diamonds (Fig. 2C), respectively, together with theoretical predictions (10) for two gaps (lines). As was implied by the raw data, the smaller gap of the $\pi$-band persists up to the bulk $T_c$, further confirming the bulk origin. While the gap of the $\pi$-band shows similar temperature dependence to the theoretical prediction, the gap of the $\sigma$-band deviates for higher measured temperatures. This implies larger interband scattering due to impurity and/or imperfection of the crystal in the surface region. But this does not affect our main conclusion, since the lowest temperature gap values are indeed different. The reduced gap parameter $2\Delta/k_BT_c$, where $k_B$ is the Boltzmann's constant, are 3.54 for the $\sigma$-band and 1.42 for the $\pi$-band, indicating weak coupling of $\sigma$-electrons most likely with the in-plane $E_{2g}$ phonon mode (24) and anomalously weak coupling of $\pi$-electrons with the same mode. The present $\sigma$- and $\pi$-band-gap values agree well with the previous angle-integrated PES results, including the temperature dependence of the gap values (25). Being compared with recent tunnelling studies (26-29), while the gap value of the $\sigma$-band is slightly smaller than the largest tunnelling gap values ($2\Delta/k_BT_c$ =4.12-4.24), the gap value of the $\pi$-band is nearly consistent with the smaller tunnelling gap values ($2\Delta/k_BT_c$=1.28-2.26).

The present ARPES studies identify the larger gap on the $\sigma$-orbital derived band and the smaller gap on the $\pi$-orbital derived band, and thus provide firm and direct experimental evidence for the two-band superconductivity (5,10) in $MgB_2$. This is inconsistent with the novel two-band scenario (7) reporting that the nesting of $\pi$-FS sheets induces a large superconducting instabilities originating in coulomb interaction and leads to a larger gap on $\pi$-FS than on $\sigma$-FS. The present results also emphasize the importance of FS sheet-dependent gap, not anisotropic gap (8). The experimental confirmation of the two-band superconductivity in MgB$_2$ gives strong support for the *ab-initio* band calculation for simple systems, and also implies that experimental efforts in combination with first principle calculations (30) will succeed to discover new materials with higher $T_c$ based on the two-band superconductivity.

<received> Style tag for received and accepted dates (omit if these are unknown).

**Acknowledgements**: We thank Prof. T. Tohyama for valuable discussion. We also thank Prof. A. Chainani for critical reading of our manuscript. This work was supported by Grant-in-aid from the Ministry of Education, Science, and Culture of Japan.

Correspondence and requests for materials should be addressed to T.Y. (e-mail: yokoya@issp.u-tokyo.ac.jp).

Figure legends

Figure 1. **A**, Intersections of calculated FS sheets (14) at $k_z = 0$ and $k_F$'s from ARPES study of single crystal $MgB_2$ (0001). Smaller two circles centered at the $\Gamma$(A) point are the 2-D cylindrical FS sheets derived from boron σ-orbital and the larger circle centered at the $\Gamma$ point are the 3D tubular FS sheets derived from boron π-orbitals. **B**, The ARPES intensity plot with respect to binding energy and $k$ along line 1 of Fig. 1**A** (the high symmetry line Σ) compared with the calculated dispersions. **C** and **D**, The same as 1**B** but measured along lines 2 and 3 of Fig.1 **A**.

Figure 2. **A** and **B**, temperature-dependent high-resolution spectra measured for the FS sheets related to the boron σ- and π-orbital derived bands, respectively. Inset shows an energy-enlarged spectra (symbols) for selective temperatures, as compared with the fitting results (lines). Here we employ higher resolution (5.0 meV for σ-band and 3.9 meV for π-band) and smaller step size (0.9 meV) to detect spectral changes as a function of temperature. To make up for the very low count rate due to using higher resolution and small sample size, the spectra shown are the sum of the APRES spectra obtained along the lines 2 and 3 containing $k_F$, and, thus, are corresponding to angle-integrated spectra but for each individual bands. This justifies the modified BCS

function analysis originally for angel-integrated spectra. For the calculations, we first simulated the normal state spectrum by assuming a linear spectral function multiplied by the FD function of 40 K and convolved with a Gaussian of full width at half maximum for the known energy resolution. The obtained spectral function is then multiplied with a modified BCS function (22) including thermal broadening (0.3-1.2 meV) and with the FD function of corresponding temperatures, and further convolved with a Gaussian to reproduce the superconducting state spectra. **C**, Plot of the superconducting gap size $\Delta(T)$ used for fitting the superconducting state spectrum. Open circles and diamonds donating the gap on the σ-derived and π-derived bands, as compared with theoretical predictions (10) (lines).

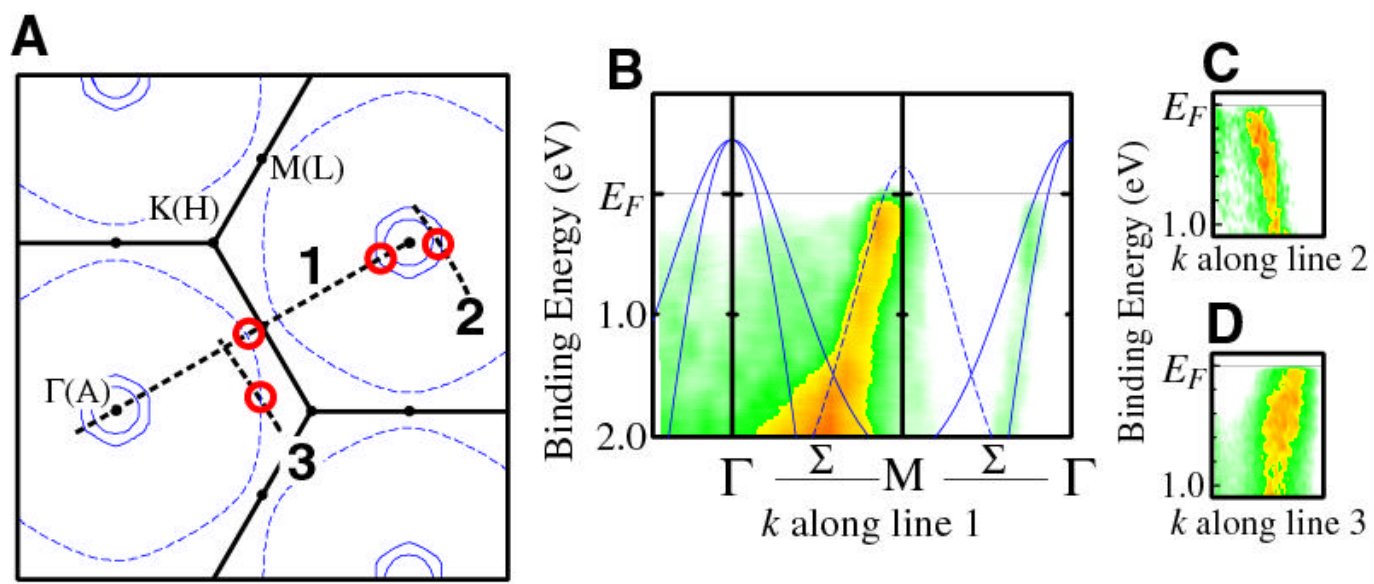

**Fig.1**

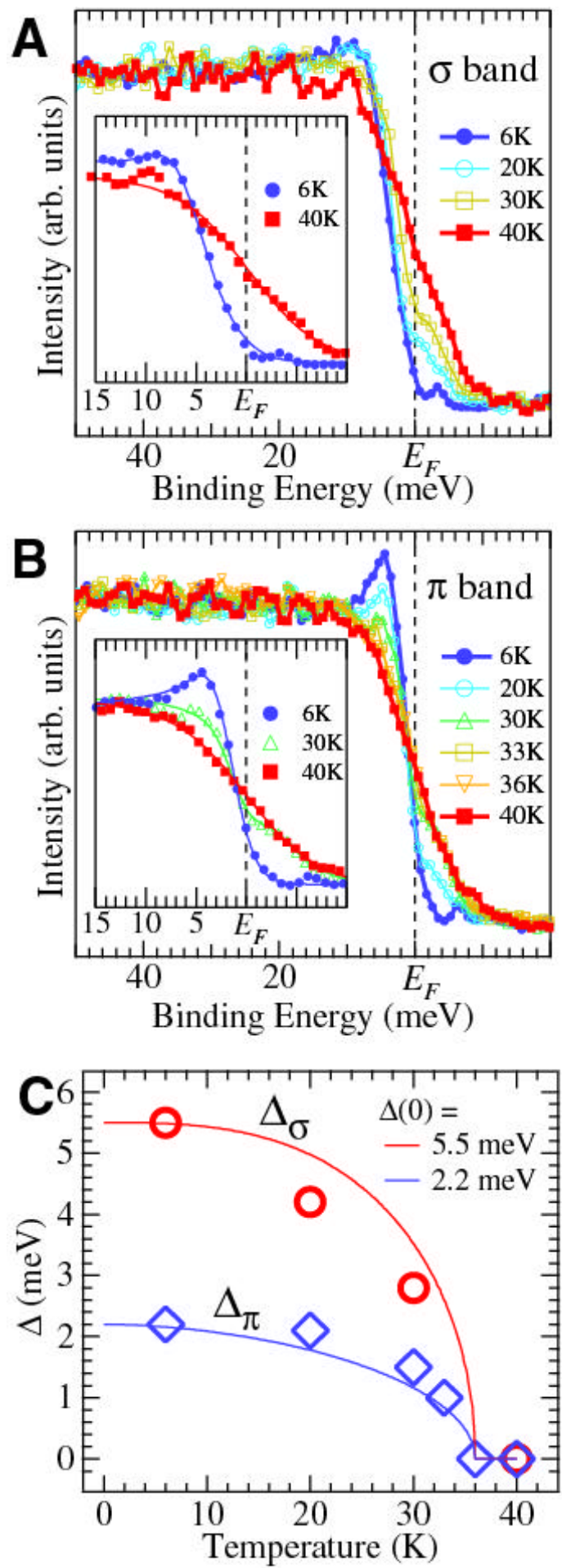

**Fig.2**